\documentclass[11pt,a4paper]{article}
\usepackage{fullpage}
\usepackage {color}
\usepackage[pdftex]{graphicx}
\usepackage[latin1]{inputenc}
\usepackage{amsmath}
\usepackage{amsfonts}
\usepackage{amssymb}

\def\lsim{\mathrel{\lower4pt\hbox{$\sim$}} 
\hskip-9.5pt\raise1.6pt\hbox{$<$}\;} 

\def\gsim{\mathrel{\lower4pt\hbox{$\sim$}} 
\hskip-9.5pt\raise1.6pt\hbox{$>$}\;}

\title{Is the Lorentz limiting speed equal to the Speed of Light? 
\\Photons {\em vs} Neutrino tests}

\date{}
\begin{document}

\maketitle

\begin{center}
{\large Jean-Marie Fr\`ere}\footnote{frere@ulb.ac.be},
{\large Simon Mollet}\footnote{smollet@ulb.ac.be} and
{\large Michel H.G. Tytgat}\footnote{mtytgat@ulb.ac.be}\\
\vskip .7cm
Service de Physique Th\'eorique,\\
Universit\'e Libre de Bruxelles, 1050 Brussels, Belgium
\end{center}
\vskip 0.5cm

\newcommand{\beq}{\begin{equation}}
\newcommand{\eeq}{\end{equation}}

\begin{abstract}

We discuss the possibility that the ``limiting speed" $c_{\mbox{\rm\tiny limit}}$ ($c_l$) appearing in Lorentz equations
 might be different ({\em i.e.}, slightly larger) than the observed speed of light $c_n$.
 We show that such a possibility can be tested by state-of-the-art  Michelson-Morley experiments,
 but also by careful measurement of neutrino speeds. It would indeed suffice to show that $c_n< c_{\nu}\leq c_{l}$.
 Quite interestingly, current limits from both approaches are competitive, in some circumstances.
 We also comment on competing tests using gamma-ray burst, assuming a dispersive character for
 the propagation of light.
\end{abstract}

\bigskip

\section{Introduction}

Lorentz transformations differ from Galilean ones by the introduction of a limiting speed,
which modifies, among other things, the law of addition of velocities.
Maxwell's equations tell that this speed is the speed of light (in vacuum).
It is however interesting to test independently the latter assertion.
Even in relatively tame contexts, the parameter limiting speed ($c_{\mbox{\rm\tiny limit}} = c_l$) in Lorentz transformations
could differ from the measured speed of light, for instance a slight mass for the photon,
or the presence of some medium with a refractive index. Notice that such a medium differs from the old-fashioned "aether" context, in that it is a physical background
within a Lorentz covariant context. To avoid confusion, we will
use $c_n$ (or $c_l/n$) to represent the measured speed of light in our surroundings
(this is an implicit reference to a possible medium playing the role of a more elaborate medium).
Such medium is by no means excluded (for instance, we have strong evidence for the presence
of dark matter in galaxies; another pretty open possibility is dark energy).
Quite obviously, the best particles to test the value of $c_l$ are (i) the photon itself, via
Michelson-Morley type experiments (but with a sensitivity much better than for the
sole rejection of Galilean transformations), and (ii) the other light particles, namely the neutrinos.
For instance,
 if one measures $c_n<c_{\nu} $, which is possible in this context ({\em i.e.} neutrinos may interact differently with the background), it implies $c_n < c_{neutrino}\leq c_{limit}$.

While interest in this possibility has obviously been revived by the sensational (but ultimately wrong \cite{Agafonova:2012wi}) claim  by
the OPERA  \cite{Adam:2011zb} collaboration of  ``neutrinos faster than light'', the present work does not
 rely in any way on that measurement; instead, it compares the merit of different methods (we will also be brought to mention different tests based on gamma-ray bursts) to test for the relation between $c_n$ and $c_{l}$.
 We have for instance shown ({\em in tempore non suspecto})  in a recent presentation \cite{Frere:2012pk} that  the OPERA claim  was in direct contradiction (although not by much) with Michelson-Morley
constraints in the framework we consider here. 
More recent results from the ICARUS experiment, operating in a similar context as OPERA, have 
put a strong bound on the difference between neutrino and photon speed, $\delta t\vert_{\mbox{\rm \tiny ICARUS}} = (0.18 \pm 0.69 (stat.) \pm 2.17 (syst.)) \, ns$ \cite{Antonello:2012be},  bringing  the neutrino speed
in line with the  current  Michelson-Morley constraints.\footnote{These measurements have been reported in August 2012 and are consistent with the former results released by the ICARUS collaboration in March 2012, which had larger statistics and systematics uncertainties, $\delta t\vert_{\mbox{\rm \tiny ICARUS}} = (-0.3 \pm 4.9 (stat.) \pm 9.0 (syst.)) \, ns$ \cite{Antonello:2012be}.}

However we want to stress the interest
for more precise checks of this fundamental issue, in what could be an interesting emulation between
neutrino timing and Michelson-Morley experiments.
In the course of the paper, we will also be brought to discuss frequency-dependent (dispersive) refraction index $n$.
This opens the way to further comparison (this time more model-specific) between astrophysical tests based
on distant sources and the Michelson-Morley ones.

Let us remark in passing that a value of $c_l$ in Lorentz transformations very slightly higher than the speed of
light (which we study here), would allow faster neutrinos without entering in direct conflict with some of the
major observations  of special relativity (dilatation of time, survival of atmospheric muons, not to mention causality,
now defined in terms of $c_l$). Our approach is thus rather conservative compared to other recent works on faster-than-light neutrinos (see, for instance, Refs.\cite{Dvali:2011mn,Alexandre:2011bu,AmelinoCamelia:2012zza}).  For convenience,  we will use the locution ``slow photons" to describe the possibility
that the measured photon speed could be less than Lorentz's limiting speed.

\section{Slow photons}

The neutrino experiments rest in fact on an indirect comparison of the neutrino speed (measured through a timing which uses the speed of the electromagnetic radiation) and the distance (also determined by photon-based measurements, including the GPS signals, of frequency within the GHz range). The simplest possibility which springs to mind, and which would bring the minimal modification of Special  Relativity, is that the effect is not due to neutrinos exceeding the limiting speed $c_l$, but to assume that for some reason, photons travel at a speed $c_n < c_l$.
By far the simplest way to achieve this is through an hypothetical photon mass (at the cost of sacrificing gauge invariance, and electric charge conservation); however existing limits on the photon mass show that any effect would be completely negligible with regard to the current observation.\footnote{For instance a result like that of  OPERA, which operates with  GPS  (frequencies in the GHz range), would have required $m_{\gamma}> 10^{-8}$ eV while the current limit is $m_{\gamma}< 10^{-18}$ eV \cite{Eidelman:2004wy}.} It would furthermore remain sensitive to the Michelson-Morley test described below.

An alternative could be a breaking of Lorentz invariance, but could equivalently be blamed on some medium in the Earth neighbourhood with an {\it ad hoc} refraction index $n$. The photon's speed in such a medium could be $c_l/n$, with $c_l$ the ``true limiting speed" \cite{Gardner:2011hg,Schechter:2012zr}. However unaesthetic such an medium may be, let us remember that other examples of such local or global structures exist, and provide a preferred frame: dark matter halos, CMB radiation, etc. Although these media are very unlikely to account for such an effect, it is a legitimate question to consider this possibility, and this, preferably, in most general terms ({\em i.e.} without referring to a specific, microscopic realization of this putative medium).

 For this purpose  we assume that  Special Relativity is preserved (with $c_n$ replaced by $c_l$), and that the Einstein velocity-addition formula can then be used to evaluate the speed of light in our reference frame. We then assume that our frame moves at a speed $v$ with respect to a medium in which the photon speed $c_n$ is defined (we set $c_n=c_l/n$). In a direction transverse to ${v}$, light velocity equal to
\begin{equation}
c_\bot = {c_l\over n}\sqrt{ 1 - {n v^2\over c_l^2}\over 1 - {v^2\over c_l^2}}
\end{equation}
 while in a direction parallel to ${v}$ we have
\begin{eqnarray} \label{averagespeed}
c_{\|}^{+}&=&\frac{\frac{c_l}{n}+v}{1+\frac{v}{n c_l}}
\label{eq:cPlusPar}\\
\nonumber\\
c_{\|}^{-}&=&\frac{\frac{c_l}{n}-v}{1-\frac{v}{n c_l}},
\label{eq:cMinusPar}
\end{eqnarray}
where $+$ and $-$ stand for the speed along or against $v$  \cite{VonLauePauli}. This is of course very standard.  Nevertheless it is of interest to see how this scenario confronts with state-of-the-art measurements of the anisotropy of the  speed of light.

In recent experiments of Michelson-Morley type \cite{Herrmann:2009zz},\cite{Eisele:2009zz}, it has been shown that the local maximal anisotropy  can not exceed
\begin{equation}\label{limit}
    (\Delta c/c)_{exp} \sim 1 \times 10^{-17}.
\end{equation}
We emphasize that this represents an improvement of about one order of magnitude compared to previous  measurements (see for instance \cite{Stanwix:2006jb}).
In these experiments the parallel ``average" velocity is measured $$c_{\|}=\frac{\text{``path length"}}{\text{``round trip time"}}=\frac{2L}{L/c_{\|}^{+}+L/c_{\|}^{-}}$$ since the experiment is based on a forth and back movement of light in a resonator. This average velocity $c_{\|}$ is
\begin{eqnarray}
c_{\|}=\left(\frac{\frac{c_l}{n}-\frac{n v^2}{c_l}}{1-\frac{v^2}{c_l^2}}\right). \nonumber
\end{eqnarray}
The theoretic resultant anisotropy which has to be compared with $(\Delta c/c)_{exp}$ is
\begin{eqnarray}
\frac{\Delta c}{c}=\frac{c_{\bot}-c_{\|}}{c_{\bot}}=\frac{\sqrt{1 - v^2/c_l^2} - \sqrt{1 - v^2 n^2/c_l^2}}{\sqrt{1 - v^2/c_l^2}}\approx \frac{v^2}{2 c_l^2}\,(n^2-1). \nonumber
\end{eqnarray}
where the last identity holds provided $v^2/c_l^2 \ll 1$.
 
 In writing this expression, we have also assumed {that the refraction index $n$ is frequency independent, or non-dispersive. We will briefly address the modifications implied by dispersive refraction index in section 5, and will briefly comment on the merit of our proposal with respect to other cosmological tests of the speed of light. Suffice it  to say here that taking a constant refractive index is the simplest example to implement the possibility that photons travel at speed lower  than $c_l$. We also emphasize that our approach is precisely equivalent (albeit with a different notation) to the formulation of contemporary tests of Lorentz invariance in terms of effective operators, the so-called Standard Model Extension \cite{Kostelecky:2002hh}.

With this in mind, we may write $n=1+\delta$ with $\delta \ll 1$ a constant ($\delta$ is called $\kappa_{tr}$ in the framework of Ref.\cite{Kostelecky:2002hh}). Then we obtain, at first order in $\delta$,
\begin{eqnarray}
\label{delta}
\frac{\Delta c}{c} \simeq  \frac{v^2}{c_l^2}\delta \simeq  \frac{v^2}{c^2}\delta\;.
\end{eqnarray}
As expected, the effect we are considering is second order in the velocity $v$.\footnote{Second order in $v/c$ comes from the compensation of the first order effects in the ``mean" velocity.}

\section{Comparison with  experimental limits}

Now we may compare the above  limit (\ref{limit}) on $\Delta c/c$ to the current constraints from the neutrino experiments on $\delta$.\footnote{One should  of course remember that the distance measurements in OPERA or ICARUS largely depend upon GPS measurements, at at frequency between 1 and 2 GHz, while the most recent Michelson-Morley experiment is performed with near-infrared photons $\sim 280$ THz. However a direct comparison of the speed of light at  these wavelengths should be comparatively easy (we do not refer here to the ionospheric propagation effects of GPS signals, which are already taken into account by the experiment).}
For this, a choice must still be made: which relative speed $v$ should we take? We will further discuss this question later,
but, to illustrate our point, for the moment we just take a rather ``conservative'' approach, associating $v$ to the rotation speed of the Earth on its
axis at the experiment point.\footnote{Clearly the other ``natural''  choices of frame  lead to much stronger exclusions. Modern laboratory tests of Lorentz invariance are usually presented within  the framework of the so-called minimal Standard Model Extension \cite{Kostelecky:2002hh} and adopt a reference frame for the speed of light that is centered on the Sun, so that  $v$ is (essentially) the Earth orbital velocity $v_\oplus/c \approx 10^{-4}$. In this case the quoted limit on $\delta$ (noted $\kappa_{tr}$ and corresponding to $\Delta c/c \lesssim 10^{-16}$  \cite{Baynes:2011nw,Hohensee:2010an}) is $\delta \lesssim 7\cdot 10^{-9}$.}

The latest null result from ICARUS \cite{Antonello:2012be} gives a bound on $\delta$:
\beq
\label{delta_icarus}
\left.\delta\right\vert_{\mbox{\rm \tiny ICARUS}}  = (0.7 \pm 2.8 (stat.) \pm 8.9 (sys.))\cdot 10^{-7}
\eeq
With $v=465 \cos 53^{o}$ m$\cdot$s$^{-1}$, where $465$ m$\cdot$s$^{-1}$ ({\em i.e.} $v/c = 9.3\cdot  10^{-7}$) is the equatorial rotation speed of the Earth, and the cosine takes into account the Berlin latitude where the Michelson-Morley experiments took place, the upper limit of (\ref{delta_icarus}) translates to the following $2\sigma$ bound on the anisotropy of the speed of light (statistical and systematic errors are added in quadrature)
\beq
\label{Delta_v_icarus}
\left.\frac{\Delta c}{c}\right\vert_{\mbox{\rm \tiny ICARUS}} \lesssim 1.4\cdot 10^{-18} < 10^{-17}
\eeq
which, remarkably, is smaller than the best limit set by current Michelson-Morley experiments.
Furthermore it is quite interesting to  note that \cite{Herrmann:2009zz} predicts improved constraints, at the $10^{-20}$ level, for the near future. {We are aware that stronger constraints on a possible refraction index for photons exist in the literature, in particular based on colliders data (see \cite{Hohensee:2008xz} and references therein).  These limits generally rest on supplementary assumptions and are thus less direct than the limits based on Michelson-Morley type experiments. }\\

The constraints are summarized in the figure, were we show, in the plane $\Delta c/c$ {\em vs} $\delta$, the exclusion sets by the current Michelson-Morley experiments and that set by ICARUS. Also shown are some possible relations between one and the other, assuming different reference frames. Clearly, if the preferred frame is the one sets by the CMB (red, upper oblique line), then it is way beyond the reach of neutrino experiments.
The other extreme is to consider that the preferred frame is somewhat attached to the Earth, as we did above. This is shown in Fig.1 as a (blue, oblique) band, which, for the sake of illustration, corresponds to  $2 v_{\rm Berlin}/3 \leq v \leq 3 v_{\rm Berlin}/2$. In this scenario, clearly the neutrino experiments are slightly ahead of Michelson-Morley experiments. 
\begin{figure}
\centering
\includegraphics[height=10cm]{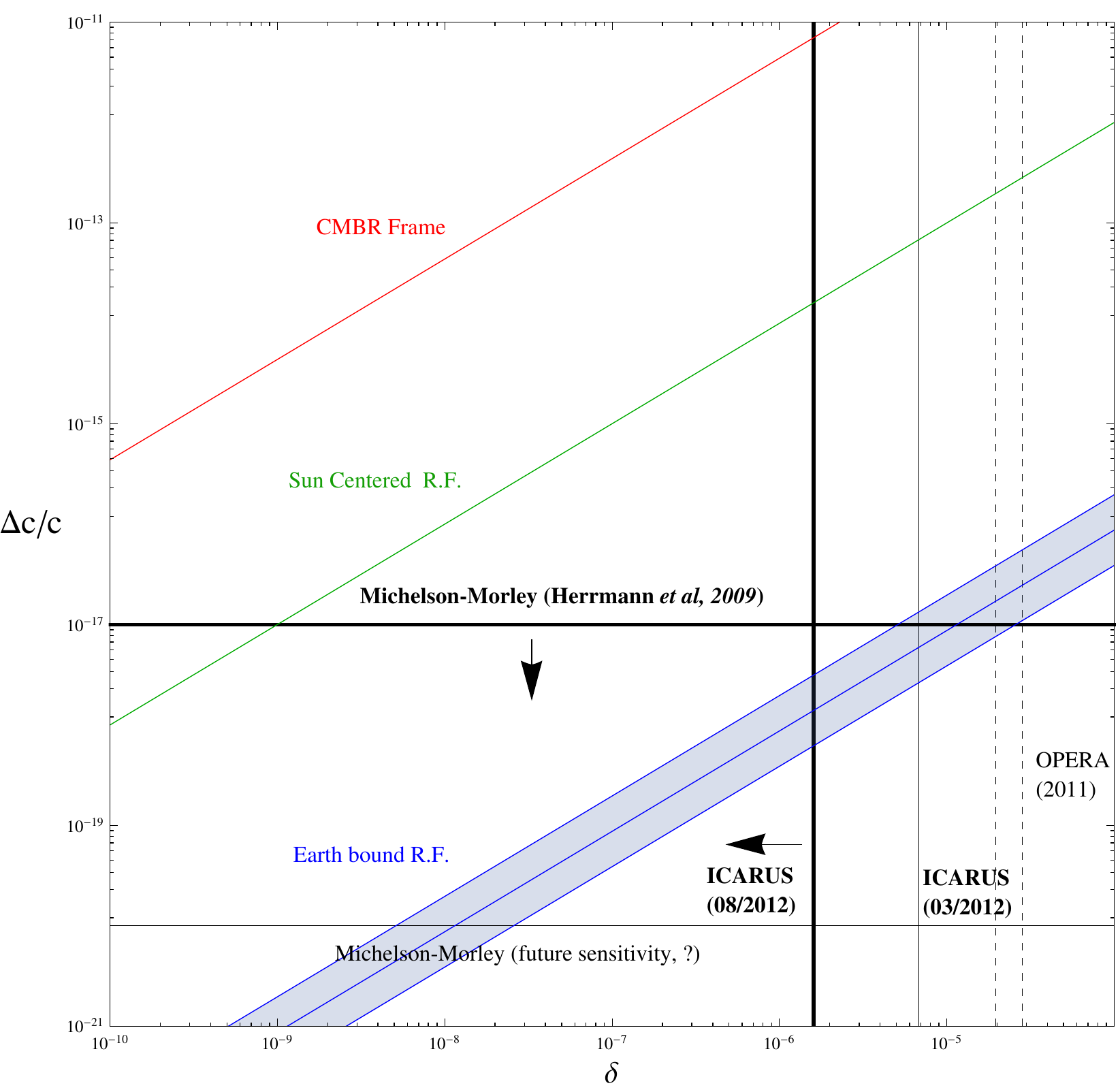}
\caption{Exclusion limits set by Michelson-Morley and neutrino experiments on ``slow photons" ({\em i.e.} Lorentz transformations limiting speed larger than c). The oblique lines refer to various possible relative speeds
which can be used in the interpretation of the Michelson-Morley result (see text).}
\end{figure}

For the sake of comparison, we show on the same plot the limit set by the former ICARUS results 
\cite{Antonello:2012hg}, and also the region that was consistent with the OPERA faster-than-light neutrinos claim \cite{Adam:2011zb},\footnote{The precise range is $\left.\delta\right \vert_{\mbox{\rm \tiny OPERA}} = (2.37\pm 0.32(stat.)^{+0.34}_{-0.24}(sys.))\cdot 10^{-5}$.}
$
\delta\vert_{\mbox{\rm \tiny OPERA}} \approx {c_l - c\over c} \approx 2.37 \cdot  10^{-5}, 
$
 which, when compared to the rotation speed of the Earth, is excluded by the Michelson-Morley experiments  (unless  the medium is in some way dragged into the Earth rotation) \cite{Frere:2012pk}.

\section{Slow photons and dark matter?}

We return now briefly  to the question of the medium approach, in relation to the choice of the relative speed $v$ in Eq.(\ref{delta}). Should it be necessary one day to resuscitate such ``aether", one obvious possibility would be to relate it to dark matter or dark energy. Clearly, as shown in the figure, we need to take the worst possible case, and thus we should envision that such medium (in particular if it is linked to some dark matter effects \cite{Frere:2007pi}) is carried along with the Earth and the Solar system in its Galactic motion, under the effect of gravitation or some hypothetical
interaction between dark and ordinary matters. This would correspond to a local over-density, distinct from the  usual  nearly-spherical galactic halo of dark matter with low angular momentum, and thus from   the dark matter velocity distributions assumed in dark matter direct detection searches (see {\em e.g.} \cite{Bertone:2010zza}). 
On the other hand, keeping a local medium static with respect to the Earth surface seems difficult without requiring some (possibly detectable) friction with ordinary matter. Thus, the rotation velocity of the Earth is a conservative choice. 
Given the potential for improvement, we expect that,  in the long run, Michelson-Morley  experiments will set the strongest constraints; nevertheless the comparison with neutrino data will remain instructive, as the tests are quite independent. 

Notice that our considerations are essentially model independent, somewhat in line with  the effective theory approach to Lorentz violation effects in the 'Standard Model Extension' framework \cite{Kostelecky:2002hh}. Although building a specific model of a dark fluid is beyond our scope, it may be of interest  to re-phrase the parameter $\delta$ in terms of microscopic properties. There are general theorems regarding the scattering of light with matter (see for instance  the refs in  \cite{Gardner:2009et}) and below we use standard notations, as for instance in  Ref.\cite{Ericson}. Basically, a constant index of refraction occurs if a fluid is composed of particles with non-zero electric polarizability, which is noted $\alpha_E$. If one considers for illustration a simple resonator of frequency $\omega_0$ in the Rayleigh scattering regime, $\omega \ll \omega_0$, then $n = 1+ \delta = 1 + {e^2\over 2 m \omega_0}\,{\rho\over m}$. The combination $\alpha_E = {e^2\over 2 m \omega_0}$ is the electric polarizability and so for more generic fluid we have  $\delta = \alpha_E {\rho/ m}$. For a specific 'particle' (atoms, nuclei, molecules,...) the standard practice is to quote the ratio $\alpha_E/m$ (in GeV$^{-1} \cdot$cm$^3$). A rule of thumb that works for simple systems is $\alpha_E \sim \alpha \times $size$^3$ and  one has typically $\alpha_E/m \sim 10^{-26}$ GeV$^{-1} \cdot$cm$^3$  for an atom. This is much smaller than what would be necessary to have a fluid with  $\delta \sim 10^{-6}$. However there exist simple systems that exhibit giant electric polarizability and for which $\alpha_E/m$ is many orders of magnitude larger than  the naive estimate (see for instance \cite{fabre}). It is possible that the same holds for dark matter, about which we know little after all. We leave this for future investigations.

\section{ Comments on a dispersive refraction index}


If the medium is dispersive the most immediate consequence is that distinction must be made between the phase and the group velocities. Writing $v_p = c_l/n$, then $v_g = c_l/n - (\omega/n)\, dn/d\omega$. Michelson-Morley experiments rest on interferometry, and so the relevant quantity for our discussion is the phase velocity. Cosmological time-of-flight constraints, as in Ref.\cite{Gardner:2009et}, are based on group velocity. The other modification is that their are further, frequency dependent, corrections to the speed of light, depending on its direction of propagation. Consider first the propagation of a photon of frequency $\omega$, as determined in the lab frame, in the direction of $v$. In Eq.(\ref{eq:cPlusPar}), the index of refraction should be given in term of $\omega^\prime$, the frequency  in the fluid frame \cite{VonLauePauli}, or
$
\omega = \omega^\prime { 1 + v n/c_l \cos\theta^\prime \over \sqrt{1- v^2/c_l^2}}
$
where $\theta^\prime$ is the angle between the velocity of the lab frame, $v$, and the direction of propagation of light as measured in the fluid frame.
To leading order  in the lab velocity $v$, we recover a result first derived by Fizeau (see {\em e.g.} \cite{VonLauePauli}),\footnote{This expression has been re-derived in \cite{Oda:2012us}, see Eq.(11), but there is a factor of $n$  missing in the numerator of the last term.  In the same work the constant term (our parameter $\delta$) is not taken into account. The differences with our results seem to stem mostly from  the limited development to order v (instead of $v^2$) in Eq.(9)
of \cite{Oda:2012us}.}
\begin{equation}
c_{\|}^{+} = {c_l\over n} + v\left (1 - {1\over n^2} + {\omega \over n} {d n\over d\omega}\right) + {\cal O}(v^2).
\end{equation}
To order $v^2$, the corresponding expression  is quite cumbersome (and not very informative), and involves the second derivative of the refraction index w.r.t. $\omega$. Here instead, we take a simple parametrization,  adapted from \cite{Gardner:2009et} and write
\begin{equation}
n(\omega^\prime) = 1 + \delta + \beta \left({\omega\over \omega^\prime}\right)^2 + \gamma \left({\omega^\prime\over \omega}\right)^2
\end{equation}
where $\omega^\prime$ ($\omega$) is the frequency of light in the fluid frame (resp. lab frame). This form may be expected on quite general grounds (it is quadratic in $\omega$  provided the fluid is not polarized) \cite{Gardner:2009et}. Furthermore one expects $\beta <0$ and $\gamma >0$, so that the group velocity is always less than $c_l$.

Using this parametrization, and taking into account the Doppler relation between $\omega$ and $\omega'$, we get, to order $v^2$ and to first order in the parameters $\delta,\beta$ and $\gamma$
\begin{equation}
{\Delta c\over c}  \approx {v^2\over c^2}(\delta + 6 \gamma)
\end{equation}
Hence, to this order, there is no contribution from the $\beta$ parameter. Notice that this result holds both for the phase velocity (used here) and the group velocity, as the difference between the two may be absorbed by  a redefinition of $\beta$ and $\gamma$. 

The implications of the above results are non-trivial. For instance, in \cite{Gardner:2009et} constraints on $\beta$ has been set using Gamma Ray Bursts (GRB), assuming that the dark fluid is composed of milli-charged dark matter particles (in this framework, the $\gamma$ parameter is negligible).  In this setup, the constraint on $\beta$, which is related to Thomson scattering and which is proportional to the dark matter density, is pretty strong,
$\vert \beta\vert_{\rm cosmic} \leq 10^{-17}$. But rescaling this to the maximum local density of dark matter at Earth \cite{Frere:2007pi}, and rescaling from GRB to  Michelson-Morley-like frequencies $\omega$, we get $\vert \beta\vert_{\rm local,M-M} \lsim 10^{-6}$, similar to our constraints on $\delta$. Our result shows that the Michelson-Morley experiments are insensitive to such corrections. Conversely, GRB constraints are insensitive to a constant contribution to the index of refraction, our parameter $\delta$ \cite{Gardner:2009et}. Hence the two approaches give complementary information regarding a possible frame dependence of the speed of light.
If the putative dark fluid has a contribution to $\gamma$ then the constraints from the body of the paper are unchanged, provided $\delta \rightarrow \delta + 6 \gamma$.

\bigskip

\section*{Acknowledgements}
This work is supported in part by the Belgian Science Policy (IAP VI/11: Fundamental Interactions), the IISN and by an ARC (``Beyond Einstein: fundamental aspects of gravitational interactions''). We thank many colleagues for discussions, in particular  Petr Tinyakov, Gaston Wilquet and Pierre Vilain, which prompted this remark.

\end{document}